\documentclass[preprint]{ptephy_v2}%
\pdfoutput = 1
\usepackage{hyperref}
\usepackage{xcolor}

\definecolor{forestgreen}{HTML}{228B22}

\newcommand{\vect}[1]{\mbox{\boldmath${#1}$}}
 
\newcommand{\be}{\begin{eqnarray}}
\newcommand{\ee}{\end{eqnarray}}

\newcommand{\beq}{\begin{equation}}
\newcommand{\eeq}{\end{equation}}
\newcommand{\beqa}{\begin{eqnarray}}
\newcommand{\eeqa}{\end{eqnarray}}

\newcommand{\lmk}{\left(}
\newcommand{\rmk}{\right)}
\newcommand{\lkk}{\left[}
\newcommand{\rkk}{\right]}
\newcommand{\lnk}{\left\{}
\newcommand{\rnk}{\right\}}
\newcommand{\LL}{\left|\left|}
\newcommand{\RR}{\right|\right|}

\newcommand{\calL}{{\cal L}}

\newcommand{\vs}{{\vect s}}
\newcommand{\vx}{{\vect x}}
\newcommand{\vy}{{\vect y}}

\newcommand{\vtx}{\tilde{\vect x}}
\newcommand{\vty}{\tilde{\vect y}}

\newcommand{\tw}{{\tilde w}}
\newcommand{\tx}{{\tilde x}}
\newcommand{\ty}{{\tilde y}}

\preprintnumber{RESCEU-18/25}
\begin{document}

\title{Nonlinear Independent Component Analysis Scheme and its application to gravitational wave data analysis}

\author[1,2,3*]{Jun'ya Kume}
\affil[1]{\small {Dipartimento di Fisica e Astronomia ``G. Galilei'', Universit\`a degli Studi di Padova, via Marzolo 8, I-35131 Padova, Italy}}
\affil[2]{INFN, Sezione di Padova, via Marzolo 8, I-35131 Padova, Italy}
\affil[3]{{\small Research Center for the Early Universe (RESCEU) and Department of Physics, Graduate School of Science, The University of Tokyo, Hongo 7-3-1, Bunkyo-ku, Tokyo 113-0033, Japan}}
\author[3]{Koh Ueno}
\author[4]{Tatsuki Washimi}
\affil[4]{Gravitational Wave Science Project (GWSP) Kamioka Branch, National Astronomical Observatory of Japan (NAOJ),
Kamioka-cho, Hida City, Gifu 506-1205, Japan}

\author[5,3,6,7]{Jun'ichi Yokoyama}
\affil[5]{Kavli Institute for the Physics and Mathematics of the Universe (Kavli IPMU), WPI, UTIAS, The University of Tokyo, Kashiwa, Chiba 277-8568, Japan}
\affil[6]{Department of Physics, Graduate School of Science, The University of Tokyo, Tokyo 113- 0033, Japan}
\affil[7]{Trans-scale Quantum Science Institute,
The University of Tokyo, Tokyo 113-0033, Japan}

\author[8]{Takaaki Yokozawa}
\affil[8]{
Institute for Cosmic Ray Research (ICRR) KAGRA Observatory, The University of Tokyo, Kamioka-cho, Hida City, Gifu 506-1205, Japan}

\author[9]{Yousuke Itoh}
\affil[9]{{\small Osaka Metropolitan University, Sumiyoshi 3-3-138, Osaka, Japan}

\email{junya.kume@unipd.it}}




\begin{abstract}%
Noise subtraction is a crucial process in gravitational wave (GW) data analysis to improve the sensitivity of interferometric detectors. While linear noise coupling has been extensively studied and successfully mitigated using methods such as Wiener filtering, subtraction of non-linearly coupled and non-stationary noise remains a significant challenge. 
In this work, we propose a novel independent component analysis (ICA)-based framework designed to address non-linear coupling in noise subtraction. Building upon previous developments, we derive a method to estimate general quadratic noise coupling while maintaining computational transparency compared to machine learning approaches. The proposed method is tested with simulated data and real GW strain data from KAGRA. 
Our results demonstrate the potential of this framework to effectively mitigate complex noise structures, providing a promising avenue for improving the sensitivity of GW detectors.
\end{abstract}

\subjectindex{xxx, xxx}

\maketitle

\section{Introduction}

Independent Component Analysis (ICA) is a signal processing technique aimed at separating observed signals into statistically independent components.~\cite{ICA1,ICA2,ICA3}.  
The most famous problem to which ICA can be used is what is called the cocktail party problem. 
A number of people a chatting in a party and their voices are monitored by a number of microphones,
which receive superposed voices of some or all of the attendees. ICA separates the sound of each source, or the voice of each participant, by leveraging the statistical independence of each source, supplemented by physical priors when available.
The separation is achieved by making use of non-Gaussianity of the sources, instead of treating it as an obstacle. In this sense, ICA occupies a unique position among methods of signal processing.

In practice, the method can be used to reduce non-Gaussian noise in an experimental output by simultaneously using the main channel and witness sensors~\cite{Morisaki:2016sxs}. 
Such a noise subtraction scheme is highly appreciable in noise-dominated experiments that needs to extract signals whose strength is much weaker than that of noise background. 
This includes, for example, ground-based gravitational wave (GW) interferometry such as LIGO, Virgo and KAGRA, which measure differential length changes between test masses with extremely high precision.
In addition to the fundamental noise sources such as laser shot noise and thermal noise, GW interferometers are exposed to various noise sources that may have interferometric or environmental origins. Complementing the efforts in noise identification and mitigation at the experimental sites, the development of noise reduction methods in data analysis is also a critical undertaking for the efficient signal extraction.

In this context, noise subtraction for the primary output of GW interferometry, referred to as GW strain data, has been extensively studied~\cite{Allen:1999wh, Driggers:2011aa, Tiwari:2015ofa, LIGOScientific:2018kdd,Davis:2018yrz,
KAGRA:2019cqm, Vajente:2019ycy, Ormiston:2020ele, Mogushi:2021deu, Yu:2021swq, KAGRA:2022frk,
Kiendrebeogo:2024qdf}. Notably, the strain sensitivity of LIGO Hanford during the O2 run was significantly improved by subtracting linearly coupled noise using Wiener filtering, aided by photodiodes that monitor beam motion and size to detect beam jitter~\cite{LIGOScientific:2018kdd}.
Meanwhile, the application of ICA has been investigated using data from the initial KAGRA (iKAGRA)~\cite{KAGRA:2019cqm} and the latest KAGRA observing run (O3GK)~\cite{KAGRA:2022frk}.
It was found that, in both time-domain and frequency-domain analyses, linearly coupled components were successfully mitigated by utilizing various auxiliary monitors.
However, the characteristics of actual noise present in data are highly diverse, requiring methods capable of addressing non-linearity and non-stationarity.

As the simplest non-linear extension, bi-linear (or quadratic non-linear) coupling is often considered in the literature for the noise characterization~\cite{hagihira2001bispectral,Hall:2024wdw} and the subtraction combining multiple witness sensors~\cite{Vajente:2019ycy}.
Ref.~\cite{Vajente:2019ycy} develops a method that subtracts sideband structures in the spectrum of data, by assuming the hierarchy between the characteristic time scale (or frequency) of different components causing the sideband.
While this assumption on the hierarchy is physically well motivated, exploring methods that go beyond such assumptions would provide a broader framework for noise reduction. 
In this regard, the machine learning-based subtraction method \texttt{Deepclean}~\cite{Ormiston:2020ele,Kiendrebeogo:2024qdf} holds the potential to handle various forms of nonlinear coupling and, by extension, address non-Gaussianity and non-stationarity.

The principle of ICA based on statistical independence is general enough to be applicable to non-linearly coupled systems~\cite{Morisaki:2016sxs}; meanwhile, its computations can be more transparent than those of machine learning.
In this work, we propose a new scheme of ICA designed to address non-linear coupling between sources and investigate its applicability to GW data analysis.
For the introduction of our new scheme, we closely follow the analytical derivation in Ref.~\cite{Yokoyama:2023jok} with technical details being more supplemented. 
We then apply this scheme to derive a method for estimating general quadratic noise coupling and evaluate its effectiveness using both simulated data and real KAGRA data.

The remainder of the paper is organized as follows. In Sec.~\ref{sec:ICA_theory}, we briefly review the concept of ICA with the simplest linear coupling and then discuss its generalization to the non-linear coupling. Then in Sec.~\ref{sec:bilinear}, we derive a subtraction method for general quadratic coupling and discuss its implementation. To demonstrate its usefulness and validity, we perform a simulation of end-to-end analysis in Sec.~\ref{sec:simulation} and then we report the result of application of our new method to the KAGRA data with hardware noise injection.
Section~\ref{sec:discussion} is devoted to the discussion and future prospects.

\section{Extending ICA from linear to non-linear mixing}\label{sec:ICA_theory}
\subsection{Linear Model}\label{sec:ICA_linear}

Here we first introduce the concept of ICA using a simple model where there are $n+1$ independent sources
of signal and noises, $\vs(t)=\, ^{t}(s_0(t),s_1(t),...,s_n(t))$ and observables $\vx(t)=\, ^{t}(x_0(t),x_1(t),...,x_n(t))$ 
which are interrelated by an instantaneous linear relation
\beq
  \vx(t)=A\vs(t)
\eeq
where $A$ is  assumed to be a time independent matrix.
Our ultimate goal is to reconstruct $\vs(t)$ out of observables $\vx(t)$, but it is not possible
to do so in full as we do not know each component of $A$.
What we do here to implement ICA is to try to find another set of variables $\vy(t)$ which are given by a linear transformation of $\vx(t)$, represented by a matrix $W$ as
\beq
 \vy(t)=W\vx(t)
\eeq
in such a way that each component of $\vy(t)$ is statistically independent.  
ICA can achieve this transformation if signals and noises have non-Gaussian distributions except for one Gaussian variable~\cite{ICA1,ICA2,ICA3}.  

The mutual independence of statistical variables may be judged by 
introducing a cost function $L(W)$ which represents a ``distance'' in the 
space of statistical distribution functionals.  As a way to measure such a distance, we 
adopt the Kullback-Leibler (KL) divergence \cite{KL} defined between two arbitrary
probability
distribution functions (PDFs) $p(\vy)$ and $q(\vy)$ as 
\beq
 D[p(\vy); q(\vy)]=\int p(\vy)\ln \frac{p(\vy)}{q(\vy)}d^{n+1}y
=E_p\lkk \ln \frac{p(\vy)}{q(\vy)}\rkk, 
\eeq
where $E_p[\cdot]$ denotes an expectation value with respect to a PDF $p$.

We examine the distance between the real distribution function of
statistically independent variables $\vs$,
$r(\vs)=\prod_{i=0}^n r_i[s_i(t)]$,
and a distribution of $\vy$, $p_y$, 
constructed from the observed distribution function of $\vx$ through
the linear transformation $\vy=W\vx$ as 
\beq
 p_y(\vy)\equiv ||W^{-1}||p_x(\vx),
\eeq
where $||W^{-1}||$ denotes the determinant of $W^{-1}$.
The cost function is then given by
\begin{align}
  L_r(W) &= D[p_y(\vy);r(\vy)]=E_{p_y}[\ln p_y(\vy)]-E_{p_y}[\ln r(\vy)]
   \nonumber \\
&= -\ln ||W|| + \int   p_x(\vx)\ln \lkk p_x(\vx)\rkk d^{n+1}x
- E_{p_y}[\ln r(\vy)] \nonumber \\
&= -H[x]- E_{p_y}[\ln\lmk||W|| r(\vy)\rmk] = -H[x]- E_{p_x}[\ln p(\vx,W)],  \label{56}
\end{align}
with
\begin{equation}
p(\vx,W) \equiv ||W|| r(\vy), 
\end{equation}
and
\beq
H[x]\equiv -\int   p_x(\vx)\ln \lkk p_x(\vx)\rkk d^{n+1}x.
\eeq
The PDF of $\vx$ in the last expression of (\ref{56})
has $W$ dependence because $p(\vx,W)$ is
a PDF of $\vx$ which is made out of the PDF of $\vy$ ($=\vs$ in this
particular case) through the relation $\vy=W\vx$.
The above formula shows that the matrix $W$ which minimizes the cost
function $L_r(W)$ also maximizes the log-likelihood ratio of $\vx$. 

Since we do not know $r(\vy)$ a priori, one may instead adopt an arbitrary 
mutually independent distribution $q(\vy)=\prod_{i=0}^n q_i(y_i)$
in the cost function.
Defining a PDF consisting of marginal distribution functions
\beq
 \tilde{p}_y(\vy)\equiv \prod_{i=0}^n \int p_y(y_0,y_1,...,y_n)\prod_{j \neq i}dy_j
\equiv\prod_{i=0}^n \tilde{p}_{i}(y_i),
\eeq
we find the following relation
\beq
L_q(W)=D[p_y(\vy);q(\vy)]=D[p_y(\vy);\tilde{p}_y(\vy)]+
D[\tilde{p}_y(\vy);q(\vy)]
\eeq
holds.  Since the KL divergence is known to be
positive semi-definite, a distribution that minimizes the first term
in the right-hand-side yields the desired linear transformation
$\vy=W\vx$ for which this term vanishes.  In this case the second
term gives a discrepancy due to the possible incorrect choice of $q(\vy)$.
In this sense it would be better to choose a realistic trial function
$q(\vy)$ as much as possible.

It is known in fact that even for an arbitrary choice of $q(\vy)$,
the correct $W$ gives an extremum of $L_q(W)$.
Hence, in order to estimate $w_{ij}$ (with $W = (w_{ij})$), we solve
\beq
  \frac{\partial L_q(W)}{\partial w_{ij}}=0 \label{partial}
\eeq
for an appropriate model of $q(\vy)$.
From the derivatives
\begin{align}
 d_W\ln||W|| &\equiv \ln||W+dW||-\ln||W||=\ln||{\vect 1}+dWW^{-1}|| \nonumber \\
&={\rm Tr}(dWW^{-1})=(W^{-1})_{ji}dw_{ij}, \nonumber
\end{align}
and
\begin{align}
 d_W f(\vy) &\equiv f\lmk (W+dW)\vx\rmk -f(W\vx)
=\frac{\partial f}{\partial y_i}dw_{ij}x_j, \nonumber
\end{align}
where $f(\vy)$ is general function of $\vy$, we find
\begin{align}
d_W L_q(W)=E_{p_y}\lkk -(W^{-1})_{ji}-x_j\frac{\partial}{\partial
 y_i} \ln q(\vy) \rkk dw_{ij}.  \label{dw}
\end{align}
Therefore, in order to satisfy Eq.~\eqref{partial}, $w_{ij}$ are determined so that the above expectation value vanishes for each index.
For example, in Ref.~\cite{Morisaki:2016sxs}, $w_{ij}$ enabling noise subtraction in GW experiments was derived by incorporating physical priors to model $q(y)$ and assuming a super-Gaussian distribution for non-Gaussian noise~\cite{Yamamoto:2016bxj}.
Note that the form of $w_{ij}$ in that case coincides with the well-known Wiener filter~\cite{Wiener}.

\subsection{Nonlinear extension}\label{sec:ICA_nonlinear}
Following Ref.~\cite{Yokoyama:2023jok}, we now extend the above analysis to the case where observables $\vx(t)$ and 
sources $\vs(t)$ are nonlinearly related. 
Our goal is to find a set of functions $\vy=\vy(\vx)$ such that each component of $\vy(t)$
is statistically independent.
For the moment, we assume that this relation holds at any time and the PDFs of $\vx$ and $\vy$ are related with each other by
\beq
 p_y(\vy)d^{n+1}y =p_x(\vx)d^{n+1}x=p_x(\vx(\vy))\LL\frac{\partial(\vx)}{\partial(\vy)}\RR d^{n+1}y,
 \eeq
from which we find
\beq
p_y(\vy) 
=p_x(\vx(\vy))\LL \frac{\partial(\vx)}{\partial(\vy)}\RR.
\eeq

As before, we wish to minimize the KL divergence
\beq
  L_q(\vy)=D[p_y(\vy),q(\vy)]=E_{p_y}[\ln p_y(\vy)]-E_{p_y}[\ln q(\vy)],~~~q(\vy)\equiv \prod_{k=0}^n q_k(y_k),
\eeq
with $q(\vy)$ being a mutually independent distribution. Each term is expressed as
\begin{align}
 E_{p_y}[\ln p_y(\vy)]&=\int
 \LL \frac{\partial(\vx)}{\partial(\vy)}\RR
  p_x(\vx(\vy))
 \ln\lkk \LL \frac{\partial(\vx)}{\partial(\vy)}\RR
  p_x(\vx(\vy))\rkk d^{n+1}y  \nonumber \\
  &=\int p_x(\vx(\vy))\ln\lkk\LL \frac{\partial(\vx)}{\partial(\vy)}\RR\rkk d^{n+1}x +
  \int p_x(\vx(\vy))\ln p_x(\vx(\vy)) d^{n+1}x,
  \end{align}
and
\begin{align}
 E_{p_y}[\ln q(\vy)]&=\int
 \LL \frac{\partial(\vx)}{\partial(\vy)}\RR
 p_x(\vx(\vy))\ln q(\vy)d^{n+1}y  \nonumber \\
  &=\int p_x(\vx(\vy))\ln\lkk
  \LL \frac{\partial(\vx)}{\partial(\vy)}\RR
  \rkk d^{n+1}x +
  \int p_x(\vx(\vy))\ln
  \lkk \LL\frac{\partial(\vy)}{\partial(\vx)}\RR q(\vy)\rkk d^{n+1}x,
  \end{align}
so that
\beq
   L_q(\vy(\vx))= \int p_x(\vx)\ln  p_x(\vx) d^{n+1}x - \int p_x(\vx)
   \ln\lkk \LL \frac{\partial(\vy)}{\partial(\vx)}\RR q(\vy)\rkk d^{n+1}x.  \label{lqxy}
\eeq   
We wish to minimize the second term of the right-hand-side of (\ref{lqxy}) with respect to the function 
$\vy(\vx)$.
Neglecting the first term of the right-hand-side of (\ref{lqxy}) 
hereafter, we may rewrite the minimization problem by an action principle
\beq
   L_q(\vy)= -\int d^{n+1}x\, \calL\lmk y_i(\vx), \frac{\partial}{\partial x_j} y_i(\vx)\rmk  \label{action}
   \eeq
with the Lagrangian
\beq
  \calL=p_x(\vx)\ln \lkk \LL \frac{\partial(\vy)}{\partial(\vx)} \RR
  q(\vy(\vx))\rkk.   \label{Lagrangian}
  \eeq
The action (\ref{action}) is minimized by a solution of the Euler-Lagrange equation:
\beq
  \frac{\delta  L_q(\vy)}{\delta y_k(\vx)} 
  =
 \sum_\ell \frac{d~}{dx_\ell}p_x(\vx)\frac{\partial x_\ell}{\partial y_k}
  -p_x(\vx)\frac{\partial ~}{\partial y_k}\ln q_k(y_k) =0,  \label{EL}
\eeq
which is a direct extension of (\ref{dw}).
Indeed, multiplying this by $x_j$ and integrating over $d^{n+1}x$ we find
\beq
  E_{p_x}\lkk -\frac{\partial x_j}{\partial y_k} - x_j\frac{\partial~}{\partial y_k}\ln q_k(y_k) \rkk =0,  \label{nldw}
\eeq
where the Jacobian component $\partial x_j/\partial y_k$ corresponds to $(W^{-1})_{j k}$ in the case the transformation is linear.
However, this ``linear order'' 
equation ~\eqref{nldw} fully captures the problem if and only if the Jacobian matrix is a constant matrix.  To incorporate nonlinear effects
in general, multiplication of Eq.~\eqref{EL} by higher-order terms ($x_l x_m$...) needs to be considered (see our model below).
We also note that the equations derived from Eq.~\eqref{EL} in such a way are merely necessary conditions to minimize the KL divergence.
For general non-linear mixing, KL divergence may develop degenerate minima with variables that are different from original $\vs(t)$ (see, e.g. Ref.~\cite{2023arXiv230316535H}). 
Therefore, we need to check that these conditions are sufficient to find the global minima.

Similarly to Refs.~\cite{Morisaki:2016sxs,KAGRA:2022frk}, the analysis above can be extended to the frequency space, by dealing with PDFs over the entire time span of interest and
assuming invertibility between the variables $\vx$ and
$\vy$ at all times. 
Let PDF $P_x$ be a functional of $\vx(t_*)$ at all times $t_*$ in the relevant range. Then it is related to that of $\vy(t_*)$ as
\beq
  P_x[\vx(t_*)][d^{n+1}x(t_*)]=  P_x[\vx(t_*)]\prod_\alpha d^{n+1}x(t_\alpha)
  = P_y[\vy(t_*)]\prod_\beta d^{n+1}y(t_\beta)=P_y[\vy(t_*)][d^{n+1}y(t_*)].
\eeq
In terms of Fourier transformed modes,
\beq
  \tx_i(f_\alpha)=\int x_i(t)e^{2\pi if_{\alpha}t}dt,
\eeq  
the above relation is expressed as 
\beq  
  P_y[\vty(f_*)]\prod_\beta d^{n+1}\ty(f_\beta)=P_x[\vtx(f_*)]\prod_\alpha d^{n+1}\tx(f_\alpha)=
  P_x[\vtx(f_*)]
  \prod_\alpha
  \prod_\beta 
  \LL\frac{\partial(\tilde{\vx}(f_\alpha))}{\partial(\tilde{\vy}(f_\beta))}\RR
  d^{n+1}\ty(f_\beta),
\eeq
where $f_*$ denotes all the frequencies collectively.
The KL divergence is then minimized by the solution of the following Euler-Lagrange equation:
\beq
  \sum_\beta \sum_\ell
  \frac{d~}{d\tx_\ell(f_\beta)}P_x[\vtx(f_*)]\frac{\partial \tx_\ell (f_\beta)}{\partial \ty_k(f_\alpha)}
  - P_x[\vtx(f_*)]\frac{\partial ~}{\partial \ty_k(f_\alpha)}\ln q_k[\ty_k(f_\alpha)]=0.\label{eq:EL_memory}
\eeq 
In the following section, we start from Eq.~\eqref{eq:EL_memory} to derive a subtraction method for a specific model of non-linear coupling.

\section{Noise subtraction for a bi-linear coupling model}\label{sec:bilinear}
\subsection{Model description and derivation of the coupling estimation}\label{sec:model_derive}
Let us consider a case where there are two statistically independent noises, $\tilde{w}_1(f)$ and $\tilde{w}_2(f)$
that can be measured by some sensors $x_1$ and $x_2$ as $\tx_1(f)=\tilde{w}_1(f)$ and $\tx_2(f)=\tilde{w}_2(f)$
and the strain channel $x_0$ measuring the GW signal $\tilde{h}(f;\theta)$ is affected by these two noises
nonlinearly as
\beq
  \tx_0(f) = \tilde{h}(f;\theta) + \tilde{n}(f)+\int df_1df_2 K_{12}(f_1, f_2)\tilde{w}_1(f_1)\tilde{w}_2(f_2)\delta (f-f_1-f_2),\label{eq:bilin_coupling}
\eeq
where $\tilde{n}(f)$ is residual Gaussian noise satisfying $|\tilde{n}(f)| \gg |\tilde{h}(f)|$ and $K_{12}(f_1,f_2)$ is an unknown coupling function that can be expressed with the convolution in time domain.
We note that general nonlinear coupling is expressed by so-called Volterra series~\cite{Vajente:2019ycy} and our $K_{12}(f_1,f_2)$ is identified with the second order kernel of that.
To subtract the contribution from measured components $\tw_1(f)$ and $\tw_2(f)$, we define the following ansatz:
\begin{align}
  \ty_0(f)=&\tx_0(f)-\int df^{\prime} W_{12}(f - f^{\prime}, f^{\prime})\tx_1(f - f^{\prime})\tx_2(f^{\prime}),\label{eq:jyica_sub}\\
   \ty_1(f)=&\tx_1(f), \\
     \ty_2(f)=&\tx_2(f). 
\end{align}
Multiplying the Euler-Lagrange equation~\eqref{eq:EL_memory} by $\tx_1(f_\mu)\tx_2(f_\nu)$ 
and integrating over the phase space, we find
\beq
\left\langle -\tx_2(f_\nu)\frac{\partial \tx_1(f_\mu)}{\partial \ty_k(f_\alpha)} -\tx_1(f_\mu)\frac{\partial \tx_2(f_\nu)}{\partial \ty_k(f_\alpha)} - \tx_1(f_\mu)\tx_2(f_\nu)\frac{\partial~}{\partial \ty_k(f_\alpha)}\ln q_k[\ty_k(f_\alpha)]\right\rangle=0,
\label{eq:EL_start}
\eeq
where we replace the ensemble average $E_{p_x}[\cdot]$ by a statistical 
average $\langle\cdot\rangle$.
For $k=0$ we can assume
\beq
  q_0(\tilde{y}_0(f_{\alpha}))=\frac{1}{2\pi S_n(f_\alpha)}\exp\lkk -
  \frac{|\tilde{y}_0(f_{\alpha}) - \tilde{h}(f_{\alpha};\theta)|^2}{2S_n(f_\alpha)}\rkk,
\eeq
where $S_n(f)$ is the power spectral density (PSD) of Gaussian noise $n$.
In real experiments, we do not know $\tilde{h}(f_{\alpha};\theta)$ (or $h(t,\theta)$) a priori. However, the weak contributions from GWs can be neglected when we take a statistical average in the real analysis. 
Hence, we can set $h(t,\theta)=0$ in the following. 
Then, in our simplified system, the first two terms in Eq.~\eqref{eq:EL_start} vanish and the minimization condition becomes
\beq
\begin{aligned}
&\left\langle  -\tx_1(f_\mu)\tx_2(f_\nu)\frac{\partial~}{\partial\ty_0(f_\alpha)}\ln q_0(\ty_k(f_\alpha))\right\rangle  \\
&=S^{-1}_n(f_\alpha)\left\langle \tx^*_0(f_\alpha)\tx_1(f_\mu)\tx_2(f_\nu)-\int df' W^*_{12}(f_\alpha-f', f')\tx^*_1(f_\alpha-f')\tx_1(f_\mu)\tx^*_2(f')\tx_2(f_\nu)
\right\rangle=0, 
\end{aligned}\label{eq:minimize_bilin}
\eeq
where the superscript $^*$ denotes complex conjugation.
Assuming that $\tx_1$ and $\tx_2$ are stationary independent noises, we find
\beq
  \langle \tx_i(f)\tx^*_j(f')\rangle =\langle |\tx_i(f)|^2\rangle
  \delta_{ij}T^{-1}\delta(f-f'),\label{eq:two_point}
\eeq
where $T$ is the duration of the time series data.
By taking the complex conjugate of Eq.~\eqref{eq:minimize_bilin} and using Eq.~\eqref{eq:two_point},
we find the estimation of $W_{12}$ as
\begin{align}
  W_{12}(f_1,f_2)=\frac{\langle \tx_0(f_1+f_2)\tx^*_1(f_1)\tx^*_2(f_2)\rangle}{T^{-1}\langle |\tx_1(f_1)|^2\rangle\langle |\tx_2(f_2)|^2\rangle}.\label{eq:nonlin_kernel}
\end{align}
Actually, by substituting Eq.~\eqref{eq:bilin_coupling} into the right-hand side of Eq.~\eqref{eq:nonlin_kernel} and using Eq.~\eqref{eq:two_point}, the ansatz for $\tilde{y}_0$ in Eq.~\eqref{eq:jyica_sub} becomes $\tilde{y}_0 = \tilde{h} + \tilde{n}$.
Therefore, Eq.~\eqref{eq:minimize_bilin} turned out to be the necessary and sufficient condition for minimization in this model. Thus, the above nonlinear model can be solved by the method discussed in Sec.~\ref{sec:ICA_nonlinear}.

In practice, however, naive frequency domain analysis may yield acausal filtering $W_{12}(\tau_1, \tau_2)$ in time domain due to, {\it e.g.}, estimation errors. (see the discussion in~\cite{Vajente:2019ycy}).
In order to mitigate such risks, Ref.~\cite{KAGRA:2022frk} performed the subtraction of linearly coupled noise only in cases where linear coherence is significant. 
This approach ensures that the estimation of noise coupling remains physically grounded and reliable.
It should be noted that the effectiveness and consistency of such an implementation have been carefully demonstrated using data in which noise was mechanically injected at the experimental site~\cite{KAGRA:2022frk}.
In the present case, we can similarly refer to the quantity called bi-coherence (see, {\it e.g.}, Refs.~\cite{hagihira2001bispectral,Hall:2024wdw}) defined as
\begin{align}
  r_{012}(f_1,f_2)=
  \frac{|
  \langle \tx_0(f_1+f_2)\tx^*_1(f_1)\tx^*_2(f_2)\rangle
  |^2
  }
  {
  \langle |\tx_0(f_1 + f_2)|^2\rangle
  \langle |\tx_1(f_1)|^2\rangle
  \langle |\tx_2(f_2)|^2\rangle
  }.
  \label{eq:bi_coh}
\end{align}
This quantity is the generalization of ``linear coherence'', allowing to capture the quadratic nonlinearity due to the bi-linear coupling between $\tx_0(f_1 + f_2)$ and $\tx_1(f_1)\tx_2(f_2)$.
In fact, one can see that $r_{012}$ characterizes the magnitude of the estimated kernel function as $|W_{12}| \propto \sqrt{r_{012}}$.
As a measure of the significance of this bilinear coupling, and to ensure that the noise coupling estimation remains physically reasonable, we set a threshold value of $r_{012}$ below which $W_{12}$ is manually set to zero in the following analysis.

\subsection{Slow approximation}\label{sec:slow_approx}
As mentioned in the introduction, we refer to Ref.~\cite{Vajente:2019ycy} as a working example of the bi-linear noise subtraction, and here we compare their method with ours.
In Ref.~\cite{Vajente:2019ycy}, assuming the hierarchy between the typical time scale of variation of $x_1(t)$ and $x_2(t)$ (or equivalently $f_1 \gg f_2$ in the frequency space), the authors demonstrate that nonlinear noise, particularly sidebands, can be subtracted by applying a linear subtraction method to $x_0(t)$ and a newly defined time series $x_{\rm bi} \equiv x_1(t)x_2(t)$.
While Ref.~\cite{Vajente:2019ycy} utilizes the Laplace variables to ensure the causality of estimated kernel function in the time domain, let us proceed with the Fourier variables as in the previous section, which facilitates a direct comparison with our present approach.
In Fourier space, the subtraction can be written as
\begin{align}
    \ty_0(f) = \tx_0(f) - \frac{\langle \tx_0(f)\tx_{\rm bi}^*(f)\rangle}{\langle |\tx_{\rm bi}(f)|^2\rangle} \tx_{\rm bi}(f),\label{eq:bilin}
\end{align}
where $\ty_0(f)$ satisfies $\langle\ty_0(f)\tx_{\rm bi}^*(f)\rangle = 0$.
While this expression itself is a well-known linear method (Wiener filtering), it is here applied to the bilinear time series $x_{\rm bi}(t)$ in the Fourier domain. 
In this sense, the method can be regarded as the simplest way to partially account for the nonlinearity of the noise coupling.

In order to compare this approximated one with our method, it might be convenient to express Eq.~\eqref{eq:bilin} in the similar way as Eq.~\eqref{eq:jyica_sub}.
Since the last term in Eq.~\eqref{eq:bilin} can be expressed as $\int df'\tx_1(f - f')\tx_2(f')$, one can immediately find the following relation
\begin{align}
W_{12}(f - f', f') = W_{12}(f) &= 
   \frac{\langle \tx_0(f)\tx_{\rm bi}^*(f)\rangle}{\langle |\tx_{\rm bi}(f)|^2\rangle}.\label{eq:bilin_kernel}
\end{align}
This expression tells us that the method in Ref.~\cite{Vajente:2019ycy} can be understood as the $f'$ independent limit of our method, where the kernel is estimated as Eq.~\eqref{eq:nonlin_kernel}. 
In other words, our scheme yields the generalization of the existing method which considers the situation where the coupling function differs for each convolved mode.
We stress again that while Eq.~\eqref{eq:bilin} is only applicable when there is a hierarchy in the characteristic frequency between two components, our method is free from such a constraint.

\subsection{Subtraction of simulated noise}\label{sec:simulation}
In order to have a better understanding of our bilinear subtraction method, we simulated noise subtraction from experimental outputs with a simple nonlinear kernel function $K_{12}$. 
To this end, we implemented Eqs.~\eqref{eq:bilin_coupling} and~\eqref{eq:nonlin_kernel} into a Python code, which from the time series data, performs the kernel estimation and subtraction with the Welch's method.
As a toy model, we adopted the following noise and kernel function:
\begin{align}
\tilde{w}_1(f) &= a_1\tilde{k}_1(f)\Theta(f - f_{1,{\rm min}})\Theta( f_{1,{\rm max}} - f),\\
\tilde{w}_2(f)  &= a_2 \tilde{k}_2(f)\Theta(f - f_{2,{\rm min}})\Theta(f_{2,{\rm max}} - f),\\
K_{12}(f_1, f_2) &= \tilde{K}_{12}(f_1, f_2)
\Theta(f_1 - f_{1,{\rm min}})\Theta(f_{1,{\rm max}} - f_1)
\Theta(f_2 - f_{2,{\rm min}})\Theta(f_{2,{\rm max}} - f_2),\\
 \tilde{K}_{12}(f_1, f_2)&=A\exp\lkk -B\lmk \frac{2 f_2}{f_{2,{\rm min}} + f_{2,{\rm max}}}\rmk^2\rkk,\label{eq:toy_kernel}
\end{align}
where $\tilde{k}_{1,2}(f)$ obey some non-Gaussian distribution.
In this study, we assumed Student's t-distribution with the degrees of freedom $\nu = 3, 10$ respectively for $\tilde{k}_{1,2}(f)$.
Although in reality the components $\tilde{w}_{1,2}$ are likely to be measured along with sensing noise from the witness sensor and other sources, we assume $\tilde{x}_{1,2} = \tilde{w}_{1,2}$ for simplicity in this analysis.
For the frequency parameters, we set $\{f_{1,{\rm min}}, f_{1,{\rm max}}, f_{2,{\rm min}}, f_{2,{\rm max}}\} = \{96,104,2,4\}$Hz. 
Although not essential to our method which uses Eq.~\eqref{eq:nonlin_kernel}, setting these frequency parameters makes the comparison to the results obtained by the slow approximation based on Eq.~\eqref{eq:bilin} more reasonable, as the latter requires $f_1 \gg f_2$.

\begin{figure}[htbp]
\begin{center}
\includegraphics[width=0.7\columnwidth]{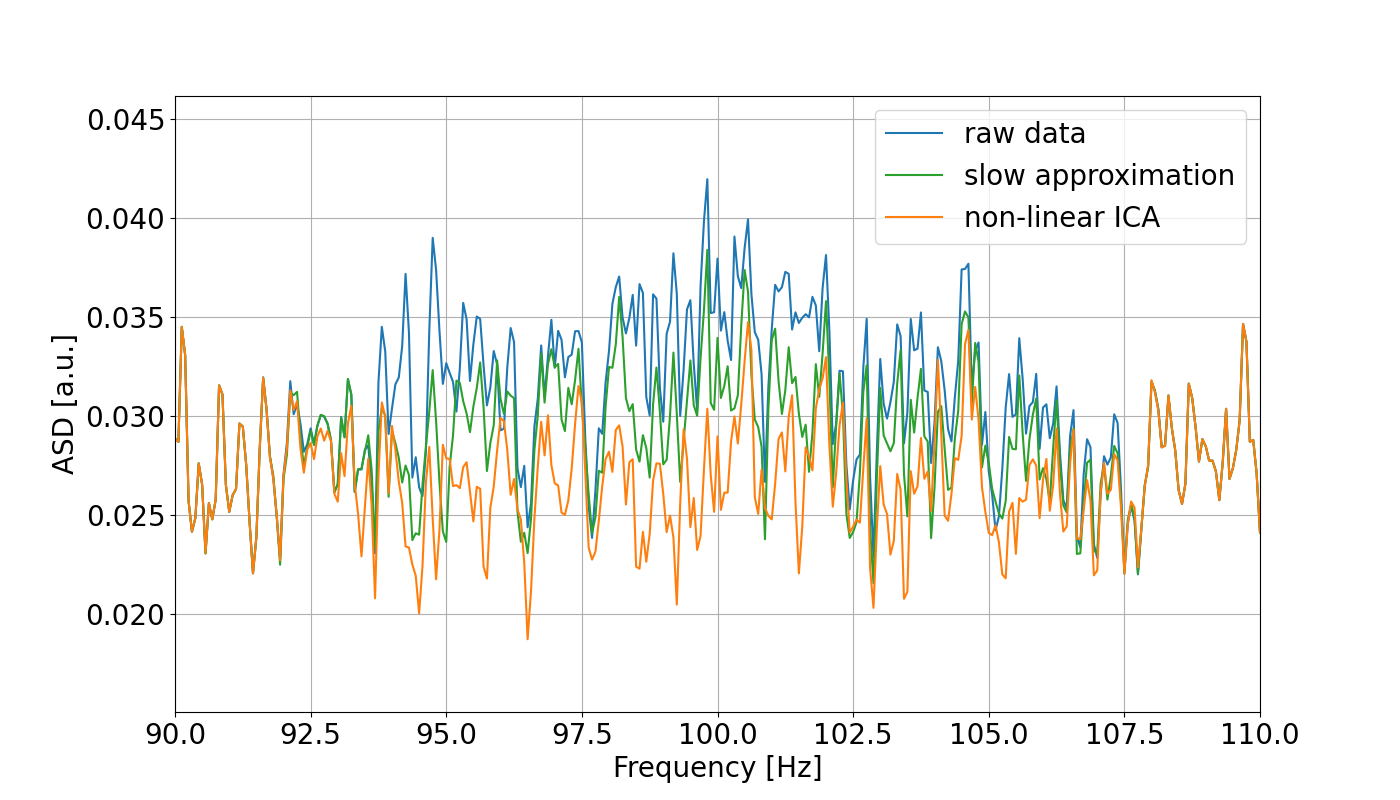}
\caption{
ASD of the data which only includes the noise components. The three curves correspond to: raw data (blue), the result after applying the slow approximation method (green), and the result after applying the non-linear ICA (orange).
}
\label{fig:noise_only_power}
\end{center}
\end{figure}

The simulated data were generated in the frequency domain and then transformed into time series. Then, we estimated the kernel function with the Welch's method. That is, we divide the entire time series data into segments with shorter duration and estimate Eq.~\eqref{eq:nonlin_kernel} by averaging over the segments. 
We assumed sampling frequency $f_s = 256$Hz and the total duration of data $T = 200$s, which is divided into segments with duration $T_s = 2$s.
We set the PSD of the Gaussian noise $\tilde{n}(f)$ in $\tilde{x}_0(f)$ as $S_n(f) = 2.8 \times 10^{-2}$.
For the remaining parameters, we assume $\{a_1, a_2, A, B\} = \{2, 1, 5\times10^2,3\}$.
In Fig.~\ref{fig:noise_only_power}, we compare the amplitude spectral density (ASD) of raw data and cleaned data. For the reference, we also show the data obtained with the slow-approximation method using Eq.~\eqref{eq:bilin}.
One can clearly see that at the frequency range $f_{1,{\rm min}} - f_{2,{\rm max}} \leq f \leq f_{1,{\rm max}} + f_{2,{\rm max}}$, the contribution from the bi-linear coupling in Eq.~\eqref{eq:bilin_coupling} is fairly subtracted and the ASD is reduced at the level of floor $\sqrt{S_n(f)}$.
We note that by construction, the slow approximation results in an incorrect estimation of $\tilde{K}_{12}(f_1, f_2)$ due to its dependence on $f_2$. 
Nevertheless, the superiority of our new subtraction method over the approximated one highlights that our method is indispensable for addressing the general frequency-dependent nature of bi-linear coupling.

\begin{figure}[htbp]
\begin{center}
\includegraphics[width=0.45\columnwidth]{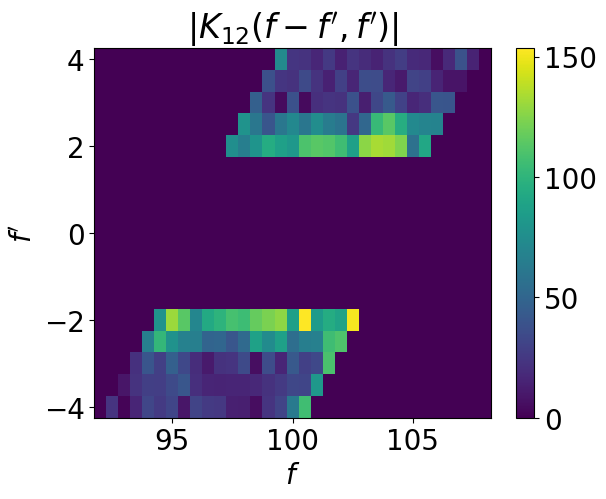}~
\includegraphics[width=0.45\columnwidth]{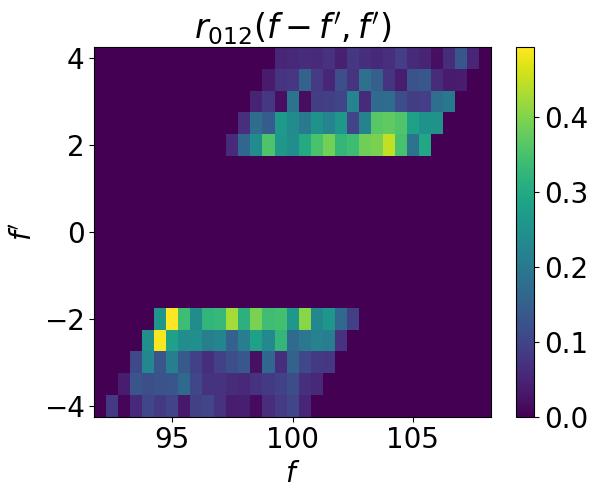}
\caption{
Left panel: Absolute value of the estimated kernel, $|K_{12}|$, which quantifies the coupling between two channels in the frequency domain.
Right panel: Bi-coherence $r_{012}$, which measures the statistical significance of the observed non-linear coupling.
}
\label{fig:noise_only_kernel_spec}
\end{center}
\end{figure}

In Fig.~\ref{fig:noise_only_kernel_spec}, the estimated kernel and bi-coherence are presented in the left and right panel, respectively.
In accordance with Eq.~\eqref{eq:toy_kernel}, we observe a decrease in the kernel function in the $f^{\prime}$ direction. 
As described below Eq.~\eqref{eq:bi_coh}, we set $K_{12}(f_1, f_2) = 0$ when bi-coherence is not large enough. From the right panel of Fig.~\ref{fig:noise_only_kernel_spec}, one can see that significant bi-coherence is estimated where coupling is large. Consequently, substantial amount of noise is effectively reduced even with a bi-coherence-based cutoff.


We then simulate analysis to detect weak signals, with an application to the GW signal search in mind.
While our frequency domain subtraction is designed to minimize the residual power, it could potentially impair the characteristics of the signal. Therefore, its practical usefulness must be carefully investigated with end-to-end analyses, including the signal detection and parameter estimation.
To this end, in the time domain, we injected a sinusoidal wave signal
\beq
h(t) = C\sin (2\pi f_c t + \phi)
\eeq
into the raw data (corresponding to the blue one in Fig.~\ref{fig:noise_only_power}). 
Then we applied matched filtering both to the raw data and to the cleaned data, on which ICA was performed after the injection.
For simplicity, we assume that only the signal frequency $f_c$ is unknown and for values of $f_c$, we compute the signal-to-noise ratio (SNR) defined as
\beq
\rho(f_c) = \frac{|z(f_c)|}{\sigma(f_c)},\label{eq:MF_def}
\eeq
where $z(f_c)$ and $\sigma(f_c)$ are expressed as
\begin{align}
 z(f_c) &\equiv 4\int df \frac{\tilde{d}^*(f) \tilde{h}(f;f_c)}{S(f)}, \\ 
 \sigma^2(f_c) &\equiv 4\int df \frac{|\tilde{h}(f;f_c)|^2}{S(f)}.
\end{align}
Here $\tilde{d}(f)$ abstractly represents Fourier variable of the raw data $x_0$ or cleaned data $y_0$ of the main channel, and $S(f)$ is the noise PSD estimated from the data $\tilde{d}(f)$. 
To smear out the effect of narrow band signal $\tilde{h}(f)$, $S(f)$ is evaluated with the running median method, where we set the number of bins involved in the estimation to be 100.

\begin{figure}[htbp]
\begin{center}
\includegraphics[width=0.45\columnwidth]{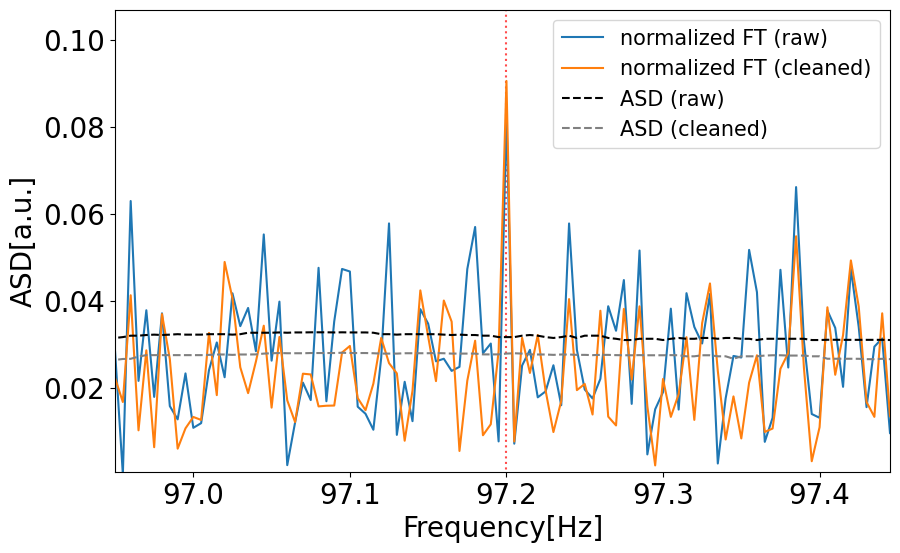}~
\includegraphics[width=0.45\columnwidth]{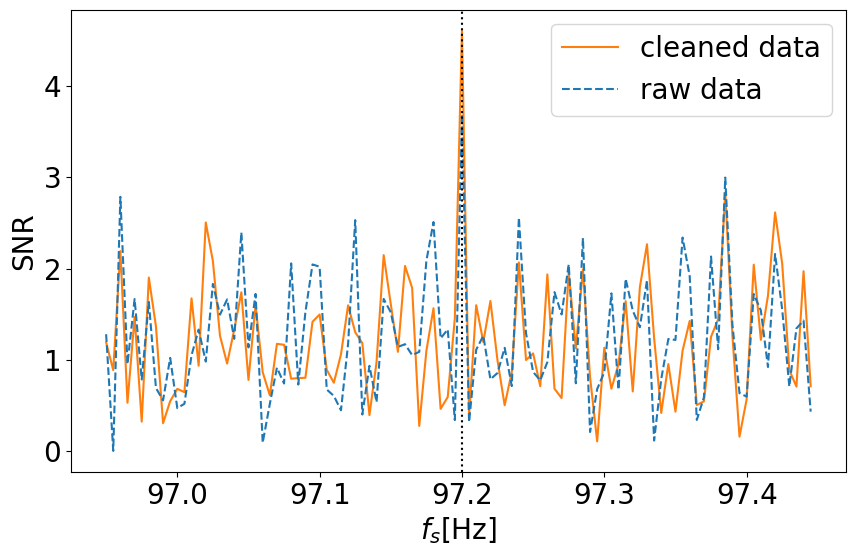}
\caption{
(Left panel) Normalized Fourier amplitude $\propto |\tilde{d}(f)|$ (solid curves) and the estimated noise ASD (dashed lines) around the injected signal frequency $f_s$, for both the raw data (blue) and the cleaned data (orange). The injected signal appears as a prominent peak in the raw data, which is increased after cleaning.
(Right panel) Matched-filter SNR for the injected signal, plotted for the raw data (dashed blue) and the cleaned data (solid orange).
}
\label{fig:sin_MF}
\end{center}
\end{figure}

Let us start with an example where the signal parameters were set as $\{C, f_c, \phi \} = \{ 9\times10^{-3}, 97.2, 3\pi/2\}$.
Notice that our subtraction method is working effectively at $f = f_c$, as seen from the suppression of the ASD of in Fig.~\ref{fig:noise_only_power}.
In the left panel of Fig.~\ref{fig:sin_MF}, (normalized) Fourier amplitude $|\tilde{d}(f)|$ and the estimated noise ASD are shown both for the raw data and the cleaned data.
In the right panel of Fig.~\ref{fig:sin_MF}, we plot $\rho(f_c)$ for the raw data and the cleaned data. One can clearly see that at the fiducial value $f_c = 97.2$Hz, SNR is increased by about 30\%. 
From the left panel, the change in the SNR can be understood in a twofold manner. 
That is, ICA (partially) removed the bi-linear component $K_{12}\tilde{w}_1\tilde{w}_2$ interfered with the signal, resulting in i)reduction in the estimated noise power $S(f)$ and ii) an increase in Fourier amplitude at $f = f_c$.
As discussed below (and also in App.~\ref{app:analytic_SNR}), the former plays important role in statistically evaluating the increase in SNR.
In contrast, the latter arises from interference between different components, which is realization dependent and manifests itself as cosine terms, as seen, {\it e.g.}, in Eq.~\eqref{eq:SNR_increase_interference}.
We note that for $f \neq f_c$, SNR after subtraction was randomly larger or smaller than that before subtraction, and $\rho$ is clearly peaked at $f = f_c$.
This indicates that subtraction did not tend to promote false positives.

However, our method is constructed to subtract any component in $\tx_0(f)$ that is coherent with $\int df^{\prime}K_{12}(f - f^{\prime}, f^{\prime})\tilde{w}_1(f- f^{\prime})\tilde{w}_2(f^{\prime})$. 
This indicates that in principle, the signal with a specific phase can also be subtracted. 
To assess the risk of signal subtraction, we examine how the outcomes of matched filtering depend on the signal phase $\phi$, using the same noise realization as in Fig.~\ref{fig:noise_only_power}.
In practice, we sample $\phi$ from a uniform distribution over $[0, 2\pi]$, inject the signal with chosen phase value, and then compare the SNR between data with and without ICA.
\begin{figure}[htbp]
\begin{center}
\includegraphics[width=0.5\columnwidth]{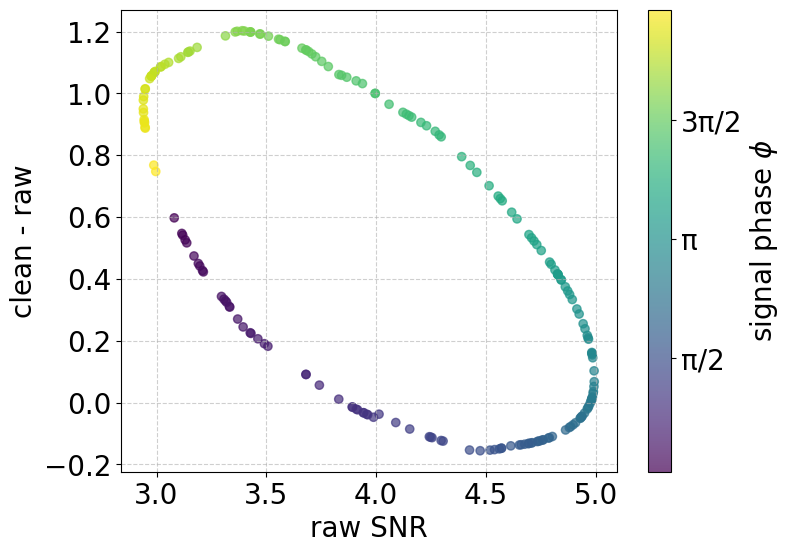}~
\includegraphics[width=0.46\columnwidth]{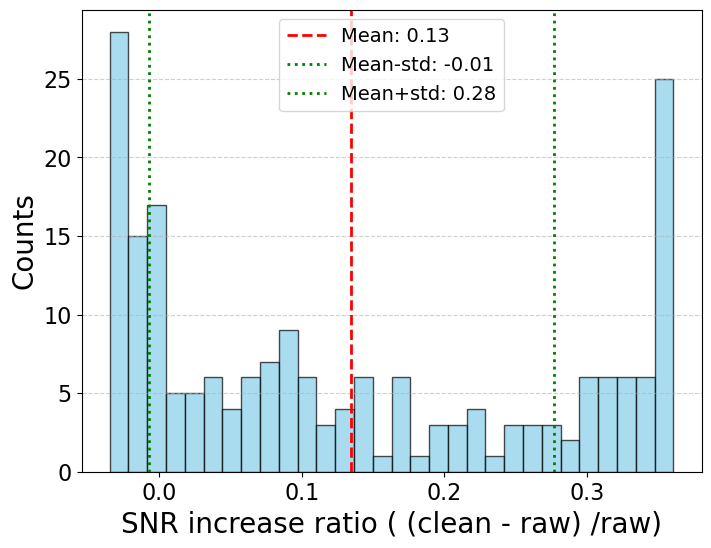}
\caption{
(Left panel) Change in SNR before (horizontal axis) and after (vertical axis) ICA, for 100 realizations of the injected signal with randomly chosen phase $\phi \in [0,2\pi]$.
Each point is color-coded by the corresponding signal phase.
(Right panel) Histogram of the fractional change in SNR across all realizations. The vertical dashed lines mark the mean, mean $\pm $ standard deviation, with the mean value showing a 
13\% improvement and the worst-case degradation being only about 
3.4\%.
}
\label{fig:SNR_sim_random_phases}
\end{center}
\end{figure}

The result is summarized in Fig.~\ref{fig:SNR_sim_random_phases}, where the left panel shows the change in SNR before and after ICA against the signal phase, and the histogram of percentage change in SNR is shown in the right panel. One can see that at the right bottom part in the left panel, there are points resulting in the reduction of SNR after ICA. Nevertheless, the right panel shows that overall, our subtraction method is beneficial to the signal detection in the present case. 
For this noise realization, we found 13\% improvement in the SNR on average. Moreover, in terms of the relative change, the decrease in SNR is only about 3.4\% even in the worst cases. 

In fact, the well-formed elliptical pattern in the left panel of Fig.~\ref{fig:SNR_sim_random_phases} indicates that the reduction in SNR after ICA is not primarily due to the (partial) subtraction of the signal itself but rather to the interference between the signal and noise components.
The right bottom part corresponds to the case where the bi-linear noise component at $f = f_c$ is coherent with the signal ({\it i.e}, $\phi_b \sim \phi$ with $\phi_b$ being the phase of the bi-linear component). 
In this case, the interference between these components additively contributes to the SNR before subtraction, indicating that subtraction of this component by ICA lead to a loss in SNR. On the other hand, they become incoherent ($\phi_b \sim \phi + \pi$) at the left upper part and in this case, interference subtractively contributes to the SNR before ICA, resulting in the larger increase of SNR after ICA.
The regions of maxima and minima appears due to the bi-linear component in this way, and as discussed in App.~\ref{app:analytic_SNR}, the interference with the unmeasured component $\tilde n(f_c)$ further shapes the overall elliptical pattern.
We expect that if signal subtraction occurs, it would be confined to a limited region around $\phi \sim \phi_b$, and significant deviations from the pattern would appear only in that area. We examined the behavior by varying the signal amplitude $C$, but significant deviation was not observed. 
Therefore, we conclude that no significant signal subtraction has occurred for this particular noise realization.

Moreover, we expect that the overall goodness (or badness) of our method is mainly determined by a balance between the strength of signal, that of bi-linear noise and unmeasured noise.
Assuming that the bi-linear component is reduced by $r_b \times 100$\% in terms of power, the average increase of SNR for our sinusoidal signal can be estimated as~\cite{Morisaki:2016sxs}
\begin{equation}
    \bar{\rho}_{\rm ICA}(f_c) - \bar{\rho}_{\rm raw}(f_c) \simeq \frac{r_bS_b(f_c)}{S_n(f_c)}\bar{\sigma}_{\rm raw}(f_c),\label{eq:SNR_increase_avg}
\end{equation}
where $S_n(f_c)$ and $S_b(f_c)$ represent PSD of unmeasured noise and bi-linear component respectively, and we here use an overbar ({\it e.g.}, $\bar{\rho}$) to represent the average over the signal phase.
This indicates that when $S_b$ (or $A$ correspondingly in our noise model in Eq.~\eqref{eq:toy_kernel}) is large, the average gain in SNR due to the subtraction also becomes large.
Indeed, we generated a separate noise dataset by doubling the value of $A$ in the kernel function and conducted a similar test with signal injection.
In this case, the SNR consistently increased due to subtraction (on average by 50\%, and at least by 37\%).
While this aligns with intuitive expectations, our method is expected to enhance signal detection and parameter estimation by improving SNR, particularly when the bi-linear component is sufficiently large with respect to the unmeasured component (which represents the intrinsic noise in the interferometer in practice).

Finally, observation made here can be extended to more complicated signals such as the chirp signal from compact binary coalescence.
For such signals, a larger number of frequency bins contribute to Eq.~\eqref{eq:MF_def}. If the power of the bi-linear component $S_b(f)$ can be effectively subtracted across all these bins, our method enhances the SNR. This requires that the bi-linear noise has a broadband contribution and that a witness sensor is available to monitor it with sufficient accuracy.

\subsection{Application to the real data with noise injection}\label{sec:real_data}
Finally, we apply our method to the real KAGRA data to figure out whether our model of non-linear coupling in Sec.~\ref{sec:model_derive} can describe a realistic situation.
We use the data taken during the commissioning period prior to the O4a observation, and  artificially inject a signal to it by actuating one of the mirrors forming the power recycling cavity. 
The actuation was applied in a specific direction (pitch) at a fixed frequency ($f = 590.1$Hz) and with constant amplitude, making the injected signal stationary and deterministic in nature. 
In the following analysis, we use a 20-minute segment of the data with injection, starting at 21:55:00 (UTC) on June 21, 2023.

The actuated mirror, referred to as the Power Recycling mirror 3 (PR3), is located at a point in the interferometer where the beam size is large, just before reaching the beam splitter. Due to this configuration, it is known that angular fluctuations of PR3 can nonlinearly couple to the interferometer’s length degrees of freedom, resulting in excess power in the corresponding signals.
In this case, the pre-existing low-frequency ($\lesssim \mathcal{O}(1)$Hz) angular fluctuations of the test mass and the injected high-frequency oscillations are expected to couple bi-linearly, leading to the appearance of sideband-like structures in the high-frequency region.
Consequently, we use the following channels in the analysis:
\begin{itemize}
    \item \texttt{K1:CAL-CS\_PROC\_PRCL\_DISPLACEMENT\_DQ} : $x_0$ (main channel)
    \item \texttt{K1:VIS-PR3\_TM\_DRIVEALIGN\_P\_OUT\_DQ} : $x_1$ (faster mode)
    \item \texttt{K1:VIS-PR3\_TM\_WIT\_P\_DQ}: $x_2$ (slower mode)
\end{itemize}
The first channel measures the displacement of power recycling cavity length (PRCL), which is one of the auxiliary length degrees of freedom in the KAGRA interferometer.
Since PRCL involves PR3 --the mirror that was artificially actuated in this study-- it is expected to exhibit the strongest response to the injection.
For this reason, PRCL was used as the primary channel in our analysis.
Although PRCL is different from DARM, the main KAGRA channel for measuring GW signals, previous observation runs have reported a significant linear coupling between DARM and PRCL. Therefore, if noise can be subtracted from the PRCL displacement channel, it would be effective for noise subtraction in DARM as well.

\begin{figure}[htbp]
\begin{center}
\includegraphics[width=0.5\columnwidth]{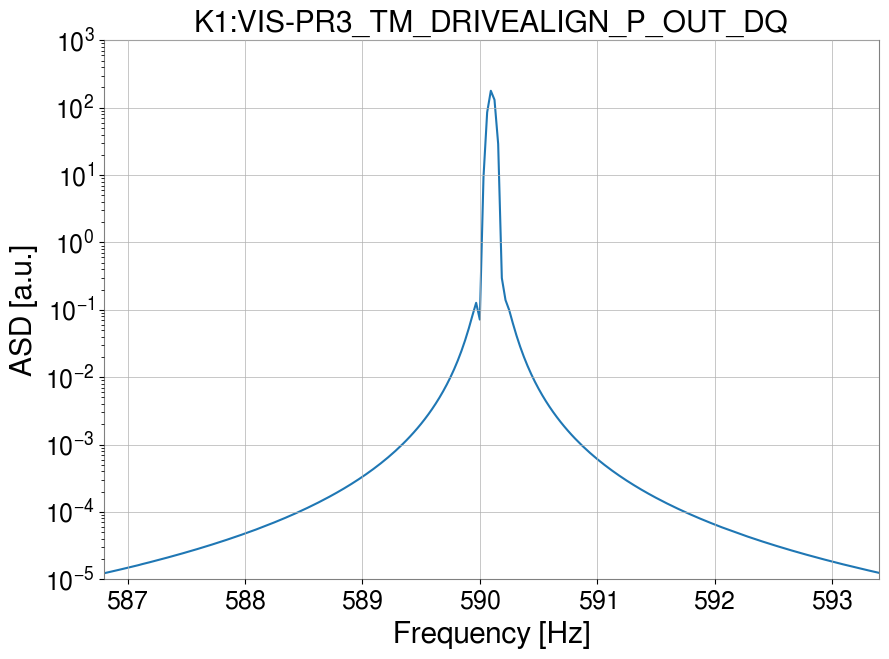}~
\includegraphics[width=0.5\columnwidth]{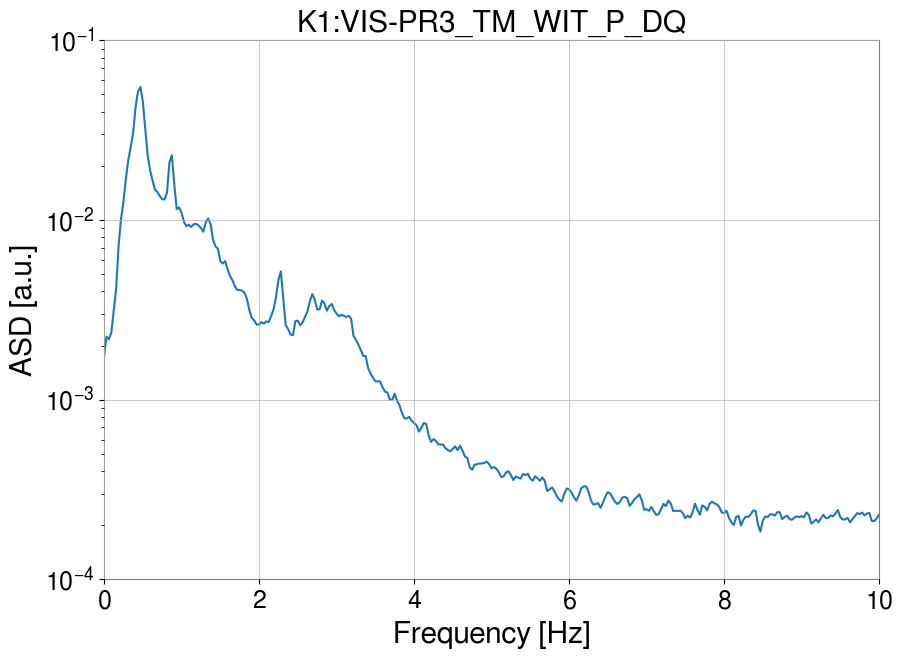}
\caption{
ASDs of witness sensors in the arbitrary unit. The left panel shows the fast component around the injected frequency $f = 590.1$Hz while the right panel presents the slow modes, which generate sidebands in the main channel through the bi-linear coupling. 
}
\label{fig:ASD_witness}
\end{center}
\end{figure}

The second channel is a witness sensor for the excitation of the PR3 mirror suspension induced by its pitch motion, and serves as a fast witness of pitch-related dynamics. 
The third channel monitors the PR3 test-mass pitch motion using an optical lever and is therefore most informative at lower frequencies.
In Fig.~\ref{fig:ASD_witness}, we plot the ASD of data from these two witness channels during injection. In the ASD of the second channel (left panel), a distinct peak appears at the injected frequency of $f=590.1$ Hz. Meanwhile, in the ASD of the third channel (right panel), multiple peaks corresponding to angular fluctuations can be observed in the low-frequency range.

\begin{figure}[htbp] 
\centering
\includegraphics[clip,width=0.8\columnwidth]{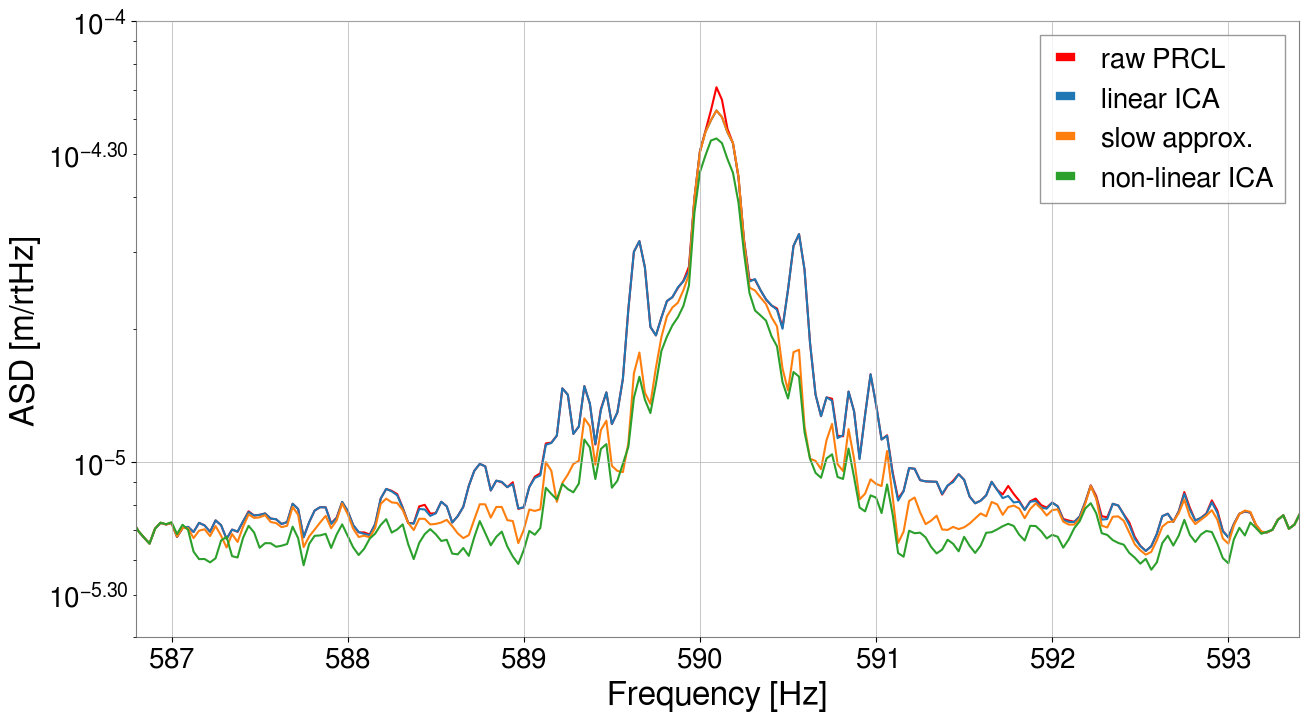}
  \caption{ASD of the PRCL around the spectral peak at 590 Hz, shown before subtraction (red) and after applying different cleaning methods: linear ICA (blue), slow-approximation method (orange), and non-linear ICA (green).
  }
    \label{fig:injection_results}
\end{figure}

In Fig.~\ref{fig:injection_results}, we plot the ASD of raw data of $x_0$ and that after noise subtraction around the injected frequency. In addition to our new subtraction method, we applied the linear subtraction using $x_1$ and the slow approximation (the one discussed in Sec.~\ref{sec:slow_approx}) for comparison.
As expected, a symmetric structure appears around the injected frequency, with two of the subpeak positions corresponding to the peak frequencies in the ASD of $x_2$. While the slow approximation (orange) also reduces these subpeaks to some extent, our method further subtracts the contributions near the central frequency as well as the floor level slightly away from the subpeaks. 

Although the additional reduction with respect to the slow approximation can appear modest in this example, such a broadband improvement is the relevant quantity for searches that integrate information over frequency and time (e.g., matched-filtering of chirp signals) as we discussed in the previous section. 
A dedicated end-to-end quantification of the detection impact (e.g., in terms of SNR or sensitive volume) depends on the target search and pipeline and is beyond the scope of this paper. 
Nevertheless, the observed reduction of both narrow features and the surrounding floor demonstrates the potential usefulness of our subtraction method on real data.
We also note that the bi-linear subtraction requires estimating three-point functions and is therefore computationally more expensive than the slow approximation. 
In practice, it is natural to apply the bi-linear subtraction after faster methods and especially when residual features remain.


The remaining subpeaks are expected to correspond to fluctuations in other angular degrees of freedom, as we found spectral peaks in those channels at frequencies different from those of $x_2$.
Therefore, integrating multiple witness sensors into our method is expected to further enhance noise subtraction performance. 
For the linear subtraction method, the information from multiple witness sensors can be combined without redundancy by performing a decorrelation through the Gram-Schmidt orthogonalization of two-point functions~\cite{KAGRA:2019cqm,KAGRA:2022frk}. 
For our bi-linear method, however, it is expected that such a decorreleation needs to be performed at the level of three-point functions.
As it is not obvious how to implement such a procedure, we leave this extension to multi-channel analysis as future work.

We finally comment on the residual around the central frequency that forms a prominent feature extending over $\sim 589$--$591\,\mathrm{Hz}$, which also includes a narrower component around $590.1\,\mathrm{Hz}$. 
One might expect that the dominant contribution in this excitation could be a standard angle-to-length coupling of the form $\delta L = x_{\rm beam}\,\delta\theta$, where $\delta\theta$ denotes, {\it e.g.}, the pitch motion of the PR3 mirror and $x_{\rm beam}$ the beam offset from the mirror center. 
Since our current witness sensors do not directly monitor fluctuations of $x_{\rm beam}$, including a dedicated beam-spot witness channel may further improve the subtraction performance.
On the other hand, if $x_{\rm beam}$ has a nonzero DC (or sufficiently slow) component, the coupling generically predicts a contribution linear in $\delta\theta$, which would be efficiently reduced by a linear subtraction based on angular witnesses. 
However, the narrower feature around $590.1\,\mathrm{Hz}$ persists even after applying the linear subtraction in the present setup. 
This suggests that the observed excitation may not be fully captured by the simple angle-to-length coupling model with the current witness set. 
Given that the underlying mechanism must account for the observed broad ($\sim 589$--$591\,\mathrm{Hz}$) structure, a more complete subtraction would likely require more advanced modeling (and/or an extended set of witness channels), which we leave for future work.

\section{Discussion}\label{sec:discussion}
In this work, we investigated the extension of ICA to the non-linearly coupled system.
Our main focus is to apply the method to the noise subtraction in data from GW experiments.
In this context, several methods~\cite{Vajente:2019ycy,Ormiston:2020ele} have been proposed to deal with quadratic nonlinear mixing. Following Ref.~\cite{Yokoyama:2023jok}, we first generalized the Euler-Lagrange equation for the KL divergence to make the scheme of ICA applicable to the non-linearly coupled system. 
Then we derived a subtraction method for the general bi-linear coupling that admits a dependence on the convolved frequency. As discussed in Sec.~\ref{sec:slow_approx}, our method can be understood as a generalization of the method proposed in Ref.~\cite{Vajente:2019ycy}, where the hierarchy between the frequency of two modes was assumed.

We then implemented our new subtraction method into a Python code, and applied it to the simulated data, where we assumed a simple toy model for bi-linear noise coupling. 
Although, by construction, the previous method cannot estimate couplings that depend on the convolved frequency, we have demonstrated that our new method effectively subtracts noise originating from the bi-linear coupling and outperforms the previous approach. This suggests that our new method serves as an effective noise subtraction approach for noise components that the existing method cannot address.
We also performed a simulation of end-to-end analysis, where a weak sinusoidal signal was injected into the data and matched filtering was performed. When the bi-linear coupling component is sufficiently large compared to the unmeasured Gaussian noise, the SNR after ICA consistently improves, with an average increase of several tens of percent. This result is consistent with the analytical calculations presented in App.~\ref{app:analytic_SNR}, indicating that our method would be beneficial to the analysis for signal detection and parameter estimation.

Finally, we applied our method to the real KAGRA data to figure out the effectiveness of general bi-linear coupling model.
To this end, we used the data from KAGRA's auxiliary length channel, where noise was mechanically injected by actuating one of the auxiliary mirrors forming the power recycling cavity. Although this was not the main experimental output sensitive to GWs, it suffices for our proof of concept study.
We found that our method not only subtracts bi-linear coupling components, such as sidebands, but also lowers the noise floor level more effectively than the previous approach. 
While it was not possible to subtract all components excited by the injection, this result suggests that our method can achieve superior noise reduction performance in real data analysis.

There are several directions in our future work. In this study, we considered only a single bi-linear coupling between two components. However, in reality, multiple components may exhibit various forms of bi-linear coupling. In the case of linear subtraction, analyses involving a large number of channels could be implemented through the orthogonalization of the two-point correlation among witness sensors~\cite{KAGRA:2019cqm,KAGRA:2022frk}. In contrast, for bi-linear subtraction, it is necessary to properly handle higher-order correlations, likely three-point correlations, {\it e.g.}, to avoid over-subtraction or addition of noise.
Towards the application to the KAGRA's observational data, we leave such an extension to multiple channels for future work.

On the other hand, the real data analysis in Sec.~\ref{sec:real_data} revealed the presence of components that cannot be subtracted using a bi-linear model. 
As the non-linearity of noise coupling has been reported in the KAGRA's observational data (see, {\it e.g.}, Ref.~\cite{Washimi:2020slk}), it is important to develop a more sophisticated nonlinear subtraction method that can address such components by leveraging our nonlinear ICA scheme.
Moreover, it would be interesting to reformulate our methodology with Laplace variable as in Ref.~\cite{Vajente:2019ycy}. From the simulation analysis conducted in this study, our practical approach of setting a cutoff in coupling estimation based on (bi-)coherence does not seem to degrade signal detection or parameter estimation. However, developing a method based on Laplace variables to construct a filter that explicitly preserves causality remains a valuable direction.

\section*{Acknowledgments}
KAGRA is supported by Ministry of Education, Culture, Sports, Science and Technology (MEXT), Japan Society for the Promotion of Science (JSPS) in Japan; National Research Foundation (NRF) and Ministry of Science and ICT (MSIT) in Korea; Academia Sinica (AS) and National Science and Technology Council (NSTC) in Taiwan.
J.K is supported by the JSPS Overseas Research Fellowships and acknowledges support from Istituto Nazionale di Fisica Nucleare (INFN) through the Theoretical Astroparticle Physics (TAsP) project. 
This work is also supported by JSPS 
Grant-in-Aid for Scientific Research (S) 20H05639, 
and the Joint Research Program of the Institute for Cosmic Ray Research (ICRR) University of Tokyo 2021-G09, 2021-G10, 2022-G09, 2022-G10, 2023-G9, 2023-G10, 2024-G08, 2024-G9. 

\appendix
\section{Analytical expressions for SNR}\label{app:analytic_SNR}
Here we supplement the analytical calculations of SNR for the sinusoidal signal to further understand the elliptical pattern in Fig.~\ref{fig:SNR_sim_random_phases}. For simplicity, we assume that the signal is not subtracted by ICA while the bi-linear component is perfectly subtracted.

Let us denote the amplitude and phase of unmeasured noise and bi-linear component at $f = f_c$ as $r_n, \phi_n, r_b, \phi_b$, respectively.
Then, by substituting 
$\tilde{d}(f_c) = r_ne^{i\phi_n}$ into Eq.~\eqref{eq:MF_def}, the SNR after ICA, $\rho_{\rm ICA}$ can be expressed as
\beq
\rho_{\rm ICA} = \sigma_{\rm ICA}\lnk 1 + \lmk \frac{\Delta_{n}}{\sigma_{\rm ICA}^2} \rmk^2 + 2\frac{\Delta_{n}}{\sigma_{\rm ICA}^2}\cos (\phi - \phi_n) \rnk^{1/2},\label{eq:rho_ICA}
\eeq
where
\begin{align}
\sigma_{\rm ICA}^2 &\equiv \frac{4|\tilde{h}(f_c)|^2}{TS_n(f_c)},\\
\Delta_{n} &\equiv \frac{4r_n|\tilde{h}(f_c)|}{TS_n(f_c)}.
\end{align}
Notice that the terms that include $\Delta_n$ are due to the interference between the signal and noise. Similarly, by substituting 
$\tilde{d}(f_c) = r_ne^{i\phi_n} + r_be^{i\phi_b}$ into Eq.~\eqref{eq:MF_def}, SNR before the subtraction can be expressed as
\beq
\begin{aligned}
\rho_{\rm raw} = \sigma_{\rm raw} &\lnk 1 + \lmk \frac{\Delta_{n}^{\prime}}{\sigma_{\rm raw}^2} \rmk^2 + 2\frac{\Delta_{n}^{\prime}}{\sigma_{\rm raw}^2}\cos (\phi - \phi_n) \right.\\
&\left.+\lmk \frac{\Delta_{b}^{\prime}}{\sigma_{\rm raw}^2} \rmk^2 + 2\frac{\Delta_{b}^{\prime}}{\sigma_{\rm raw}^2}\cos (\phi - \phi_b) + 2\frac{\Delta_{n}^{\prime}\Delta_{b}^{\prime}}{\sigma_{\rm raw}^4}\cos (\phi_n - \phi_b)\rnk^{1/2},
\end{aligned}\label{eq:rho_raw}
\eeq
where
\begin{align}
\sigma_{\rm raw} &\equiv \frac{4|\tilde{h}(f_c)|^2}{T(S_n(f_c) + S_b(f_c))},\\
\Delta_{n}^{\prime} &\equiv \frac{4r_n|\tilde{h}(f_c)|}{T(S_n(f_c) + S_b(f_c))},\\
\Delta_{b}^{\prime} &\equiv \frac{4r_b|\tilde{h}(f_c)|}{T(S_n(f_c) + S_b(f_c))}.
\end{align}
Here again $S_b$ is PSD of the bi-linear component.
From Eqs.~\eqref{eq:rho_ICA}--~\eqref{eq:rho_raw}, the difference between SNRs before and after ICA is characterized by two distinct contributions. The first and statistically important one is overall increase of $\sigma$ due to the reduction of bi-linear component in total PSD, which is expressed as Eq.~\eqref{eq:SNR_increase_avg} in the case of partial subtraction.
The other one is the subtraction of the terms in Eq.~\eqref{eq:rho_raw} related to the bi-linear component:
\beq
\delta\rho_1 \equiv \lmk \frac{\Delta_{b}^{\prime}}{\sigma_{\rm raw}^2} \rmk^2 + 2\frac{\Delta_{b}^{\prime}}{\sigma_{\rm raw}^2}\cos (\phi - \phi_b) + 2\frac{\Delta_{n}^{\prime}\Delta_{b}^{\prime}}{\sigma_{\rm raw}^4}\cos (\phi_n - \phi_b).\label{eq:SNR_increase_interference}
\eeq
The appearance of maxima and minima in the plot of the left panel in Fig.~\ref{fig:SNR_sim_random_phases} can be understood as a result of the interference from this second term.

The overall behavior of ellipse is further determined by the interference term in Eq.~\eqref{eq:rho_ICA}
\beq
\delta\rho_2 \equiv 2\frac{\Delta_{n}}{\sigma_{\rm ICA}^2}\cos (\phi - \phi_n).
\eeq
When $\phi_n \sim \phi_b + \pi$, maximization of $\delta\rho_2$, {\it i.e.}, $\rho_{\rm ICA}$ works in the direction of minimizing $\delta\rho_1$, {\it i.e.}, $\rho_{\rm ICA} - \rho_{\rm raw}$. Consequently, the ellipse is tilted downward to the right as in the left panel of Fig.~\ref{fig:SNR_sim_random_phases}. 
While not shown in this paper, we also observed another case where the ellipse is tilted upward to the right, which would correspond to the case $\phi_n \sim \phi_b$.

\bibliographystyle{ptephy}
\bibliography{ref}
\end{document}